# Detection of proton tracks with LiF Fluorescent Nuclear Track Detectors

P. Bilski[1], B. Marczewska[1], M. Sankowska[1], A. Kilian[1], J. Swakoń[1], Z. Siketić[2], P. Olko[1]

*1 - Institute of Nuclear Physics Polish Academy of Sciences (IFJ PAN), Kraków, Poland*
*2 - Ruđer Bošković Institute, Zagreb, Croatia*

Abstract
Fluorescent nuclear track detectors based on LiF crystals were successfully applied for detection of proton induced tracks. Irradiations were performed with protons with energy ranging from 1 MeV up to about 56 MeV and for all proton energies the fluorescent tracks were observed. The tracks are not continuous, but consist of a series of bright spots. The gaps between spots tend to narrow with decreasing proton energy (increasing ionization density). For the highest of the studied energies, the spots are scattered so sparsely, that it is not possible to link spots belonging to one track. The intensity (brightness) of the fluorescent tracks increases with the increasing LET and agrees well with the trend established earlier for various heavier ions.

## 1. Introduction

Fluorescent detection of nuclear tracks is a radiation measuring method originally developed by Akselrod and co-workers using $Al_2O_3$:C,Mg single crystals (Akselrod et al., 2006a; Akselrod et al., 2006b) and successfully introduced into dosimetric practice in the various fields of application (Akselrod et al., 2020; Akselrod and Kouwenberg, 2018; Akselrod and Sykora, 2011; Greilich et al., 2013; Muñoz et al., 2023; Sawakuchi et al., 2016). In the last few years another material was found to be suitable for application as a fluorescent nuclear track detector (FNTD): undoped lithium fluoride crystals (Bilski and Marczewska, 2017; Bilski et al., 2019b). The physical mechanism enabling fluorescent imaging of particle tracks in LiF is based on the creation by ionizing particles $F_2$ color centers in the crystal lattice. These centers, when excited with the blue light (wavelength around 445 nm), emit photoluminescence in the red spectral range (peaked at 670 nm). With the fluorescent microscope, using high magnification and a sensitive digital camera, it is possible to image the radiation tracks with a resolution below 1 micrometer. Track intensity, which is the intensity of the fluorescent light emitted from a track, depends on the ionization density, i.e., on the amount of the locally deposited energy. LiF crystals have been successfully used to image tracks of various ions, ranging from helium to iron (Bilski et al., 2019a). However, in the case of protons, for high-energy beams, like those used in radiotherapy, it has been so far difficult to observe the single tracks of primary protons due to the low ionization density of these particles. Preliminary analysis of proton-irradiated LiF crystals revealed the presence of some fluorescent tracks but only in the form of very scarcely distributed spots. The number of these spots was by more than an order of magnitude lower than the number of protons impinging on the crystal. Their fluorescence intensity was very low – at a similar level as the intensity of the tracks produced by gamma radiation. It was therefore difficult to decide whether the observed tracks were produced by primary protons, by protons with degraded energy, or by some secondary particles. On the other hand, it was known that low-energy protons may produce quite distinct tracks, as it happens in the LiF crystals irradiated with thermal neutrons, where the tracks produced by 2.73 MeV $^3$H nuclei (products of the nuclear reaction of neutrons with $^6$Li nuclei) are well visible (Bilski et al., 2018). Therefore, the goal of the present work was to study more closely the capability of LiF FNTDs for detecting protons of both low and high energy. The subject is relevant not only to measurements of radiotherapy proton beams, but also to dosimetry of cosmic radiation, where protons are abundant.

LiF FNTDs were recently flown to the lunar orbit in the frame of the Artemis mission (MARE experiment) (Berger, 2023) and are currently also employed in the ongoing DOSIS-3D project on board the International Space Station (Berger et al., 2016).

## 2. Experimental details

### *2.1 Samples*

Lithium fluoride single crystals were grown with the Czochralski method at the IFJ PAN. The grown bulk crystals were then cut with diamond saws into square plate samples of a typical size of about 4x4x1 mm and polished. Next, the crystal samples were heated at a temperature close to LiF melting point (820-830 °C) for 10 minutes, which improved the quality of the crystal surface by removing small scratches caused by polishing.

### *2.2 Proton irradiations*

The main proton irradiations were performed at the IFJ PAN in Kraków, exploiting the proton beam from AIC-144 cyclotron, with the initial energy of 60 MeV (Swakon et al., 2010). To obtain proton beams with lower energy, a PMMA range-shifter was applied. The dose was kept constant at the level of 20 mGy (in terms of absorbed dose in water, as measured with an ionization chamber), which corresponds to a proton fluence ranging from about $1.9\times10^6$ cm$^{-2}$ to $1.1\times10^7$ cm$^{-2}$, depending on proton energy. Positions of the detectors along the proton Bragg curve are illustrated in Fig.1.

Additionally, the experiments were supplemented by irradiations with monoenergetic low-energy protons at the Laboratory of Ion Beam Interactions of Ruđer Bošković Institute in Zagreb using a 6 MV Tandem Van de Graaff (VdG) accelerator. Irradiations were performed at the nuclear microbeam end-station, where ion beam size and current can be precisely controlled (Jakšić et al., 2007). The energy of protons ranged between 1 MeV and 6 MeV (±0.2%) and the fluence was kept below $10^5$ cm$^{-2}$. The details of the used proton beams are given in Table 1. The main experiments consisted of exposures with proton beams directed perpendicularly to the crystal surface. Additionally, some crystals were irradiated with proton beams directed in parallel to the surface. In the case of the Van de Graaff irradiations, for technical reasons the smallest achievable angle to the surface was 20°.

Table 1. The details of the proton beams exploited for irradiation of FNTDs. In the case of AIC-144 the given energy values represent mean energy calculated with the SRIM program (Ziegler et al., 2010) for a given depth in water. The depth in water was calculated from the thickness of the PMMA using an experimentally determined water equivalent ratio (WER) of 1.165. LET in LiF was calculated with the SRIM for the mean proton energies.

| | Depth in water [mm] | Energy [MeV] | Fluence [cm$^{-2}$] | LET (LiF) [keV/µm] |
|---|---|---|---|---|
| AIC-144 cyclotron | 1.47 | 56.3 | $1.06\times10^7$ | 2.42 |
| | 3.37 | 54.1 | $1.02\times10^7$ | 2.50 |
| | 14.82 | 38.7 | $1.09\times10^7$ | 3.26 |
| | 24.36 | 20.5 | $1.09\times10^7$ | 5.44 |
| | 25.52 | 17.2 | $1.11\times10^7$ | 6.26 |
| | 26.28 | 14.7 | $1.09\times10^7$ | 7.11 |
| | 26.65 | 13.4 | $1.06\times10^7$ | 7.68 |
| | 27.42 | 10.1 | $1.04\times10^7$ | 9.62 |
| | 27.88 | 7.9 | $1.06\times10^7$ | 11.64 |
| | 28.18 | 6.4 | $1.07\times10^7$ | 13.73 |
| Van de Graaff | - | 6 | $6.67\times10^4$ | 14.49 |
| | - | 5 | $6.25\times10^4$ | 16.70 |
| | - | 3 | $6.25\times10^4$ | 25.52 |
| | - | 2 | $7.14\times10^4$ | 34.40 |
| | - | 1 | $7.14\times10^4$ | 53.65 |

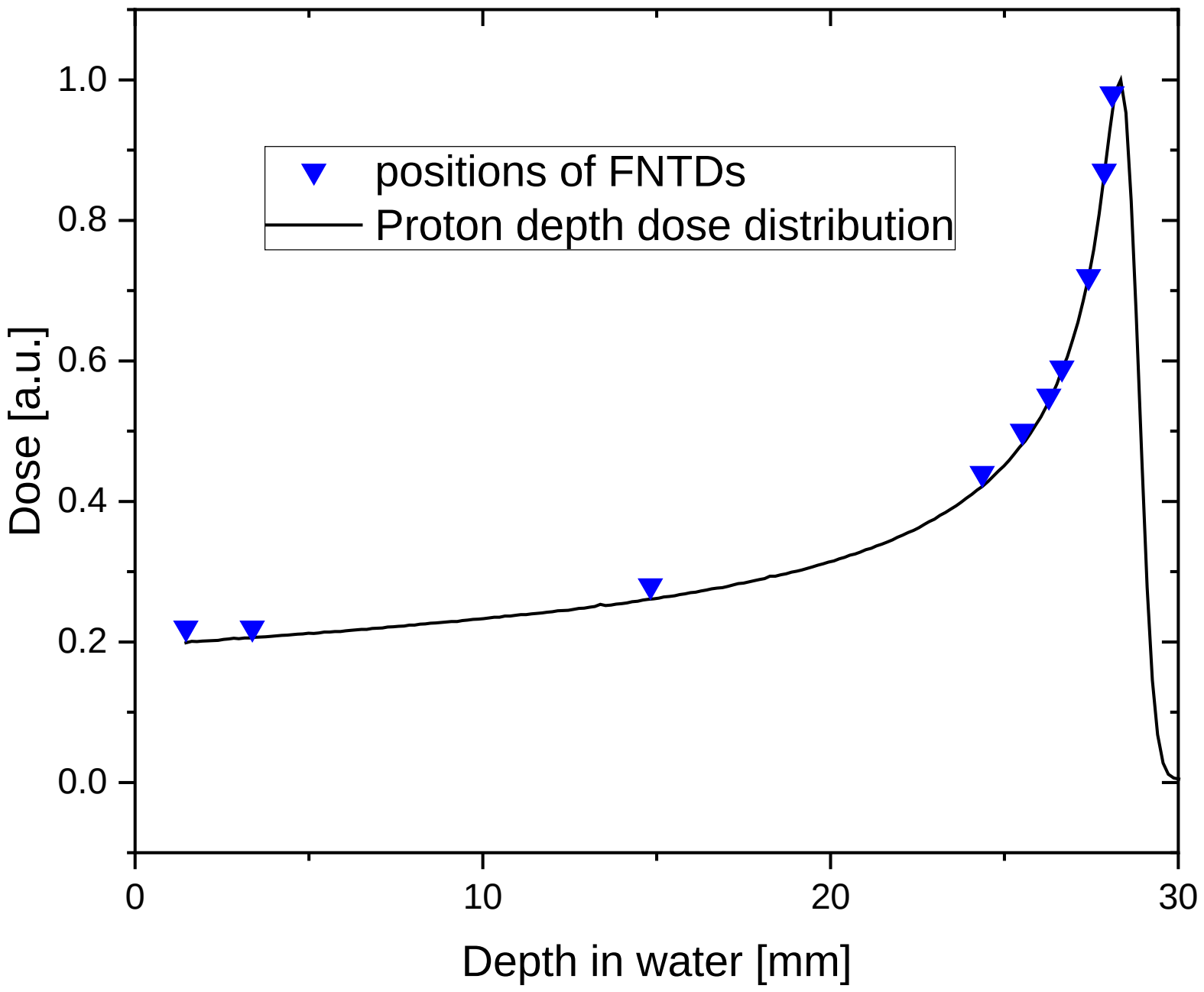

Fig. 1. Illustration of positions of LiF FNTDs during irradiations at different depths along the proton Bragg curve (proton dose rate measured with an ionisation chamber).

*2.3 Microscopic observations*

Microscopic fluorescent images of LiF crystals were acquired using a Nikon Eclipse fluorescence wide-field microscope with a DS-Qi2 camera. For excitation, the pE-100 illumination system with 440 nm LEDs was used. A band-pass filter ET445/30 was used for excitation light, while a long-pass ET570lp for emission. The observations were conducted with a 100× TU Plan ELWD (NA 0.80) objective lens. One pixel of the obtained images corresponded to about 0.07 µm. Each measurement consisted of registering a stack of image frames (slices) acquired in 1 µm steps at different depths into a crystal, starting just below the crystal surface. The 1 µm distance approximately corresponds to the vertical resolution of the used optical setup. The acquisition time of a single frame varied between 20 and 40 seconds.

*2.4 Image analysis*

Image processing and analysis were performed with the ImageJ/Fiji software (Schindelin et al., 2012). The background was numerically subtracted using the self-developed Fiji plugin. The tracks were then analyzed using the "Analyze Particles" command of Fiji, which provided data on the number of tracks (spots) and the intensity of each of them. The track intensity was defined as the maximum intensity within a bright spot divided by the time of the image acquisition. In these analyses, only tracks from the first slice were included (to avoid multiple counting tracks of the same particle), and the tracks with an area lower than 6 pixels were disregarded (to avoid counting artifacts). For qualitative analysis, the maximum intensity projections (MIP) were generated, which are a kind of two-dimensional representation of three-dimensional images that enable a fast interpretation of the track geometry. In a MIP for each pixel position in a stack of images, the brightest pixel among all layers of the stack is taken. The MIP images are schematically illustrated in Fig. 2. A circular shape of a track in the MIP means that a particle was directed perpendicularly to the crystal surface, while an elongated shape corresponds to a particle directed under an angle. Curved shapes of tracks may be also observed, indicating changes of a particle trajectory due to interactions with nuclei.

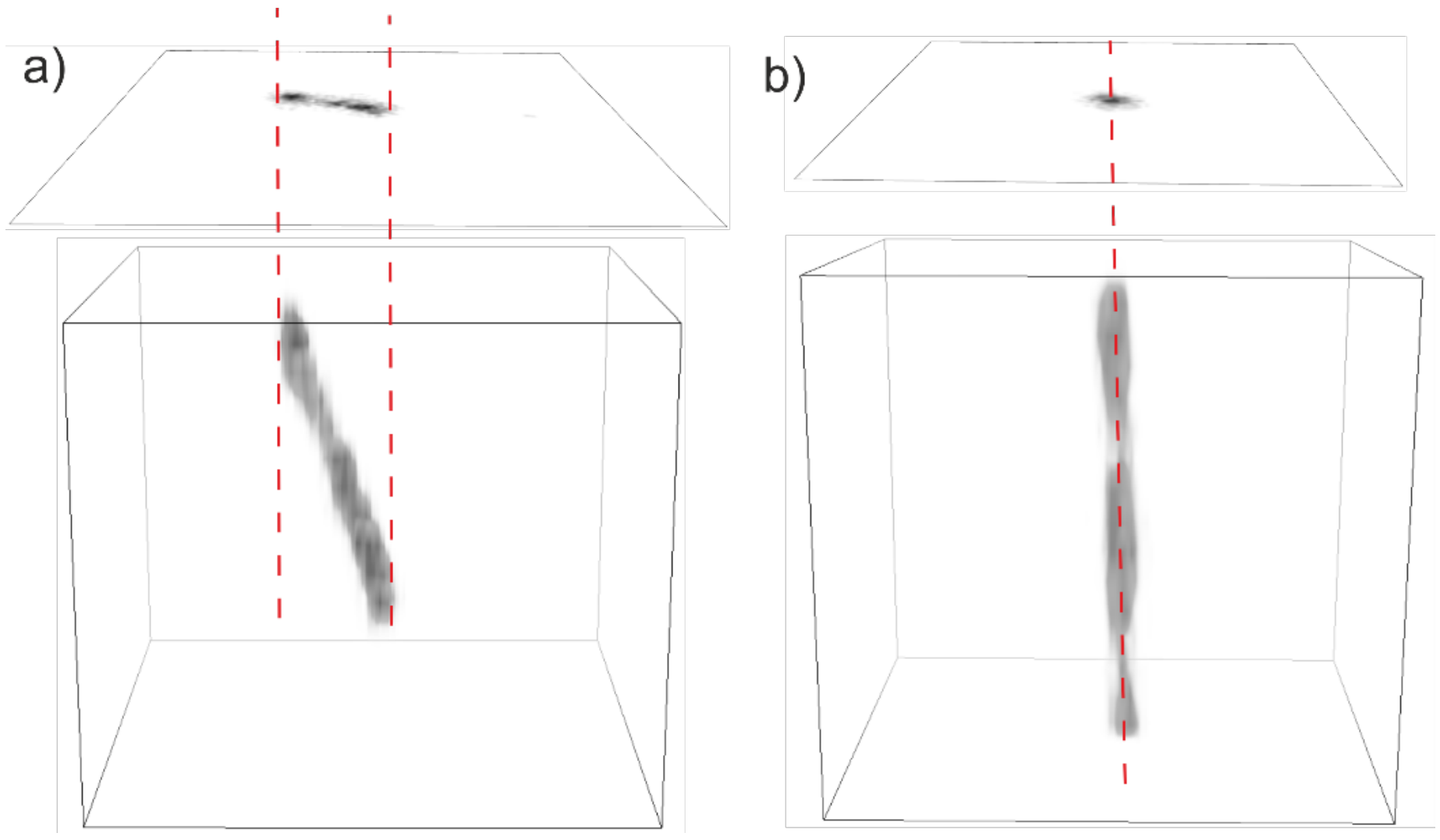

Fig. 2. Schematic illustration of the maximum intensity projections (MIP) of fluorescent tracks. a) – a particle directed under an angle to the crystal surface, b) – a particle directed perpendicularly to the crystal surface.

## 3. Results and discussion

Fig. 3 illustrates examples of fluorescent tracks produced by the protons directed nearly parallel to the crystal surface. It should be noticed that tracks are not continuous, but consist of a series of bright spots. The gaps between spots tend to narrow with decreasing proton energy (increasing ionization density). This is a result of fluctuation of fluorescence intensity along a track, which is a typical feature of fluorescent tracks and was earlier observed for ion tracks. Fig. 4 illustrates this effect for a track of an iron ion (the result of the earlier experiments at the HIMAC accelerator, Chiba, Japan (Bilski et al., 2019a). Since the ionization density for this particle is high ($LET_{LiF}$>400 keV/µm), track intensity much exceeds the background and as a result the track is continuous. However, if the signal-to-noise ratio was significantly lower (e.g. background intensity near 3000 units), only parts of the track with the peak intensity would exceed the background, which would produce a series of bright spots, exactly as occurs in the case of proton tracks. The physical mechanism of this effect seems to be related to the microscopic pattern of energy loss by a particle, but it is yet not fully understood and this subject is currently under study.

In the case of higher energy protons (fig. 5), the distance between spots exceeds 10 µm and it is hard to decide which spot belongs to which track.

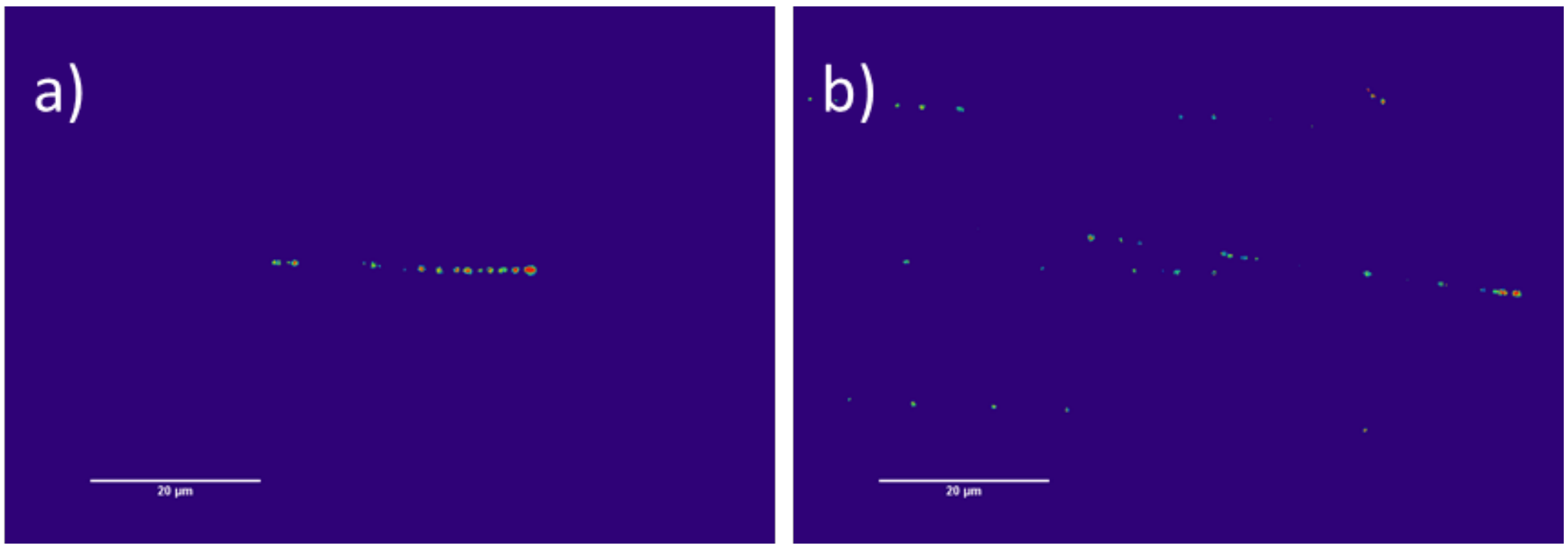

Fig. 3. Examples of track images of low-energy protons (VdG irradiations) directed nearly parallel to the crystal surface: a) 2 MeV, b) 5 MeV (energy of protons entering the crystals). Images represent maximum intensity projections of stacks of 9 frames taken in 1 µm steps. The color scale arbitrary.

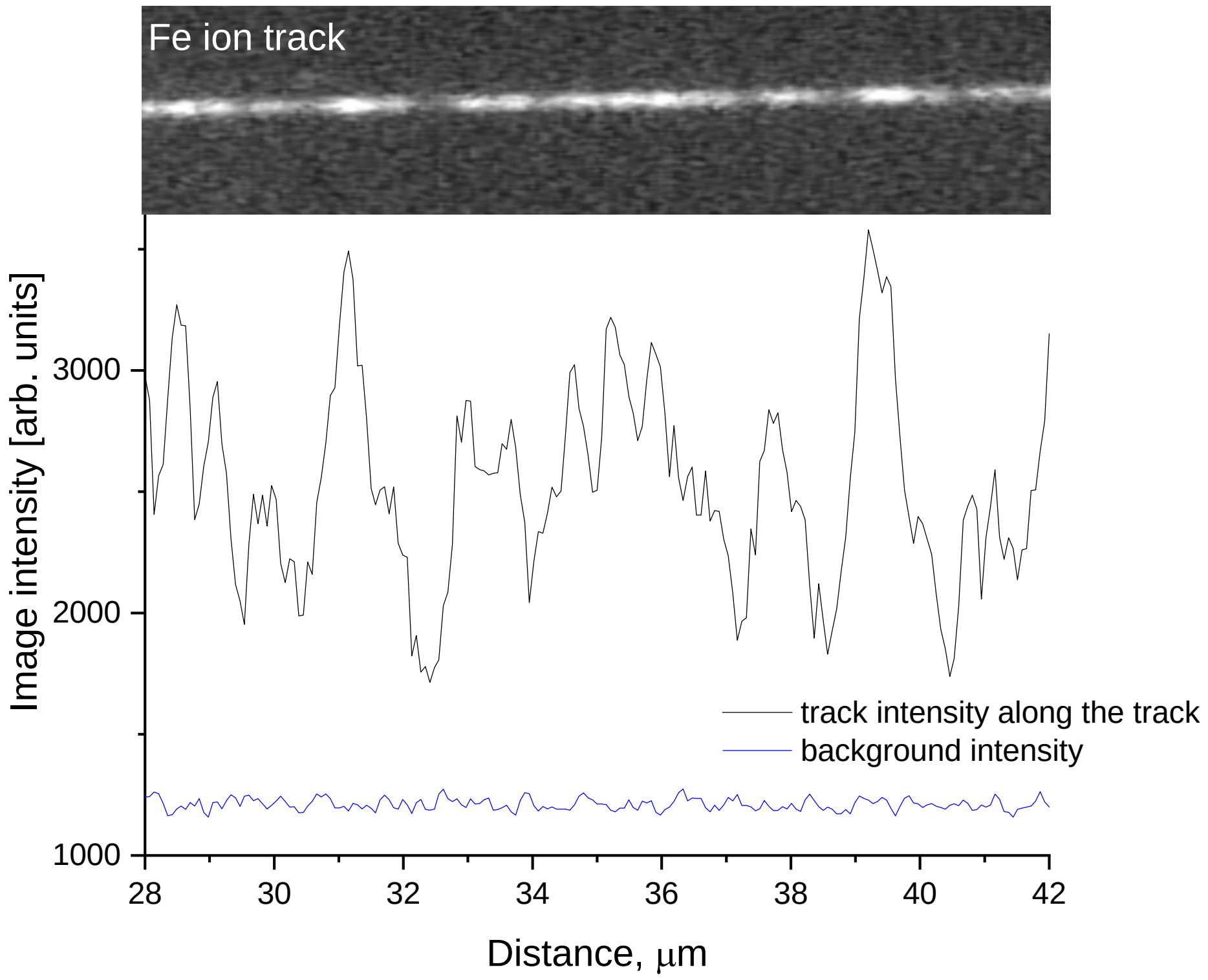

Fig. 4. Illustration of fluctuations of track intensity along a track of an iron ion (irradiations at HIMAC, Chiba) compared to the intensity of the background around the track. Inset: actual image of the track (background not subtracted).

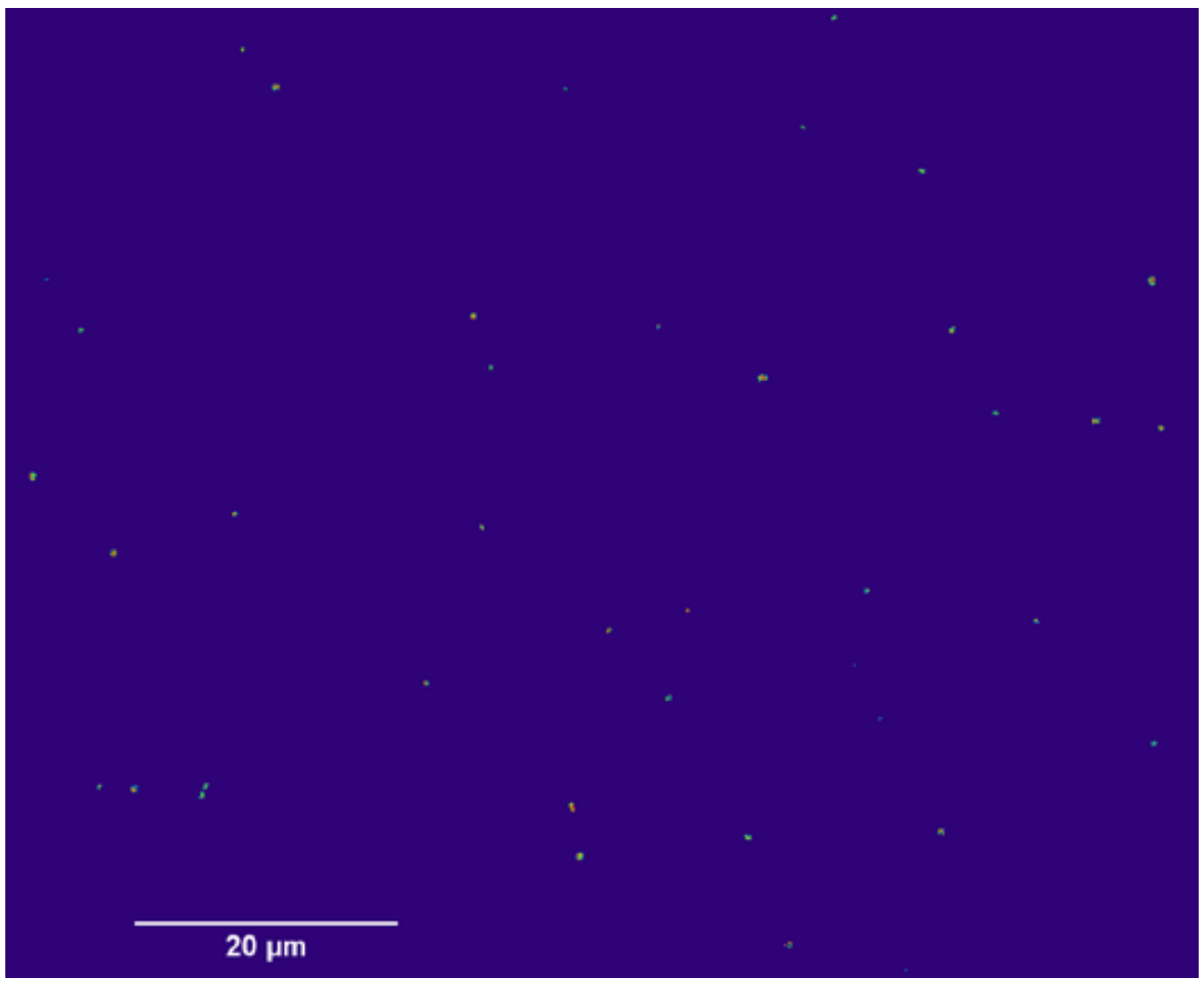

Fig. 5. Examples of tracks produced by the cyclotron proton beam directed nearly parallel to the crystal surface. Proton energy at entering crystal 56.3 MeV. The image was registered about 0.2-0.3 mm from the crystal edge (proton beam directed from the left side of the image).

Fig. 6 shows examples of tracks measured for proton beams directed perpendicularly to the crystal surface, for the highest and the lowest of the studied energies. The left column of the figure (panels a and c) presents images registered in the first 1 µm slice, while the right column (panels b and d) presents the maximum intensity projections (MIP). The number of tracks visible in the MIP images is much higher than for a single slice. The reason for this effect lies in the observed distance between bright spots along a track. For protons with high energy, these gaps between spots may extend over several micrometers. Since the thickness of a single image layer is only 1 micrometer, it is obvious that in such a layer only part of the tracks may be visible.

In the case of 56.3 MeV protons, the tracks visible in the MIP image remain nearly circular, as mostly they correspond to single spots. Oppositely, in the case of 7.9 MeV protons many tracks of elongated shape are visible, which are produced by the scattered protons or some secondary particles. In general, the brightness and the number of visible tracks increase with decreasing proton energy. The latter is mainly the effect of the mentioned decreasing length of the gaps between bright spots.

Fig. 7 presents examples of tracks registered in the first 1 µm slice after irradiations with the proton beams from the Van de Graaff accelerator. Similarly, as for higher energies, the brightness and the number of tracks increase with decreasing energy.

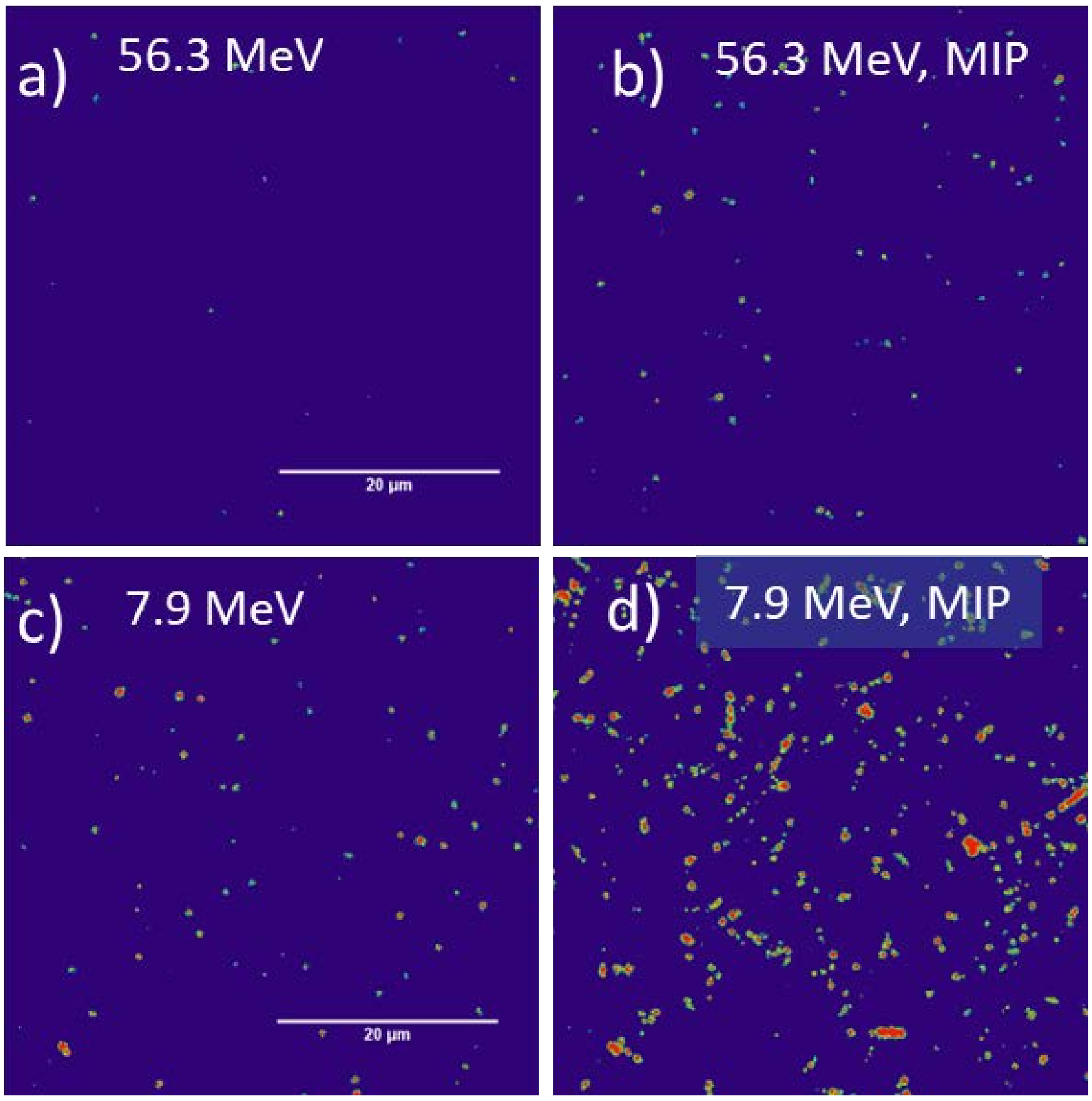

Fig. 6. Examples of track images after irradiation with the proton beam from the cyclotron: a) 56.3 MeV, the first 1 μm slice of LiF, b) 56.3 MeV, maximum intensity projection (MIP) of 20 μm layer, c) 7.9 MeV, the first 1 μm slice of LiF, d) 7.9 MeV, maximum intensity projection of 20 μm layer. The scale is the same for all images. The color scale is arbitrary.

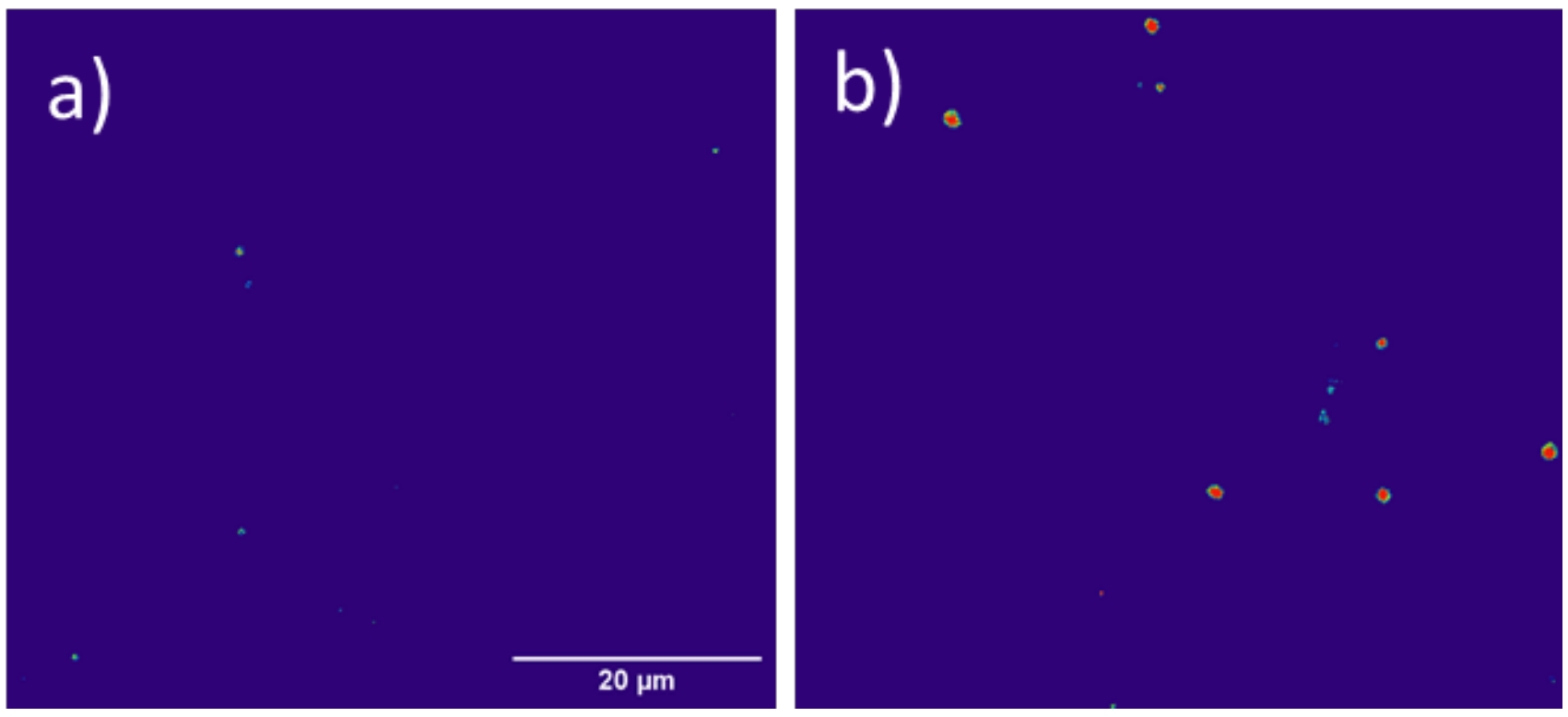

Fig.7. Examples of track images registered in the first 1 μm slice of LiF after irradiations with the proton beams from the Van de Graaff accelerator: a) 6 MeV, b) 1 MeV. The scale is the same for both images. The color scale is arbitrary.

The measured intensities of tracks for a given proton energy are not uniform but present quite a wide distribution of values. This is partly a result of the track properties (presence of gaps and possible nonuniformity of crystal properties) and partly by the actual distribution of proton energies in a beam. Fig. 8 compares track intensity distributions for three chosen energies. With decreasing energy, the distribution broadens in the high-intensity directions, which is a result of energy straggling of protons during passage through a layer of PMMA absorber. In general, the intensity of tracks increases with decreasing proton energy, in accordance with the so far observed trend for various ionizing particles.

This is presented quantitatively in Figs 9 and 10. The agreement between intensities of the fluorescent tracks induced by protons and by heavier ions is quite good.

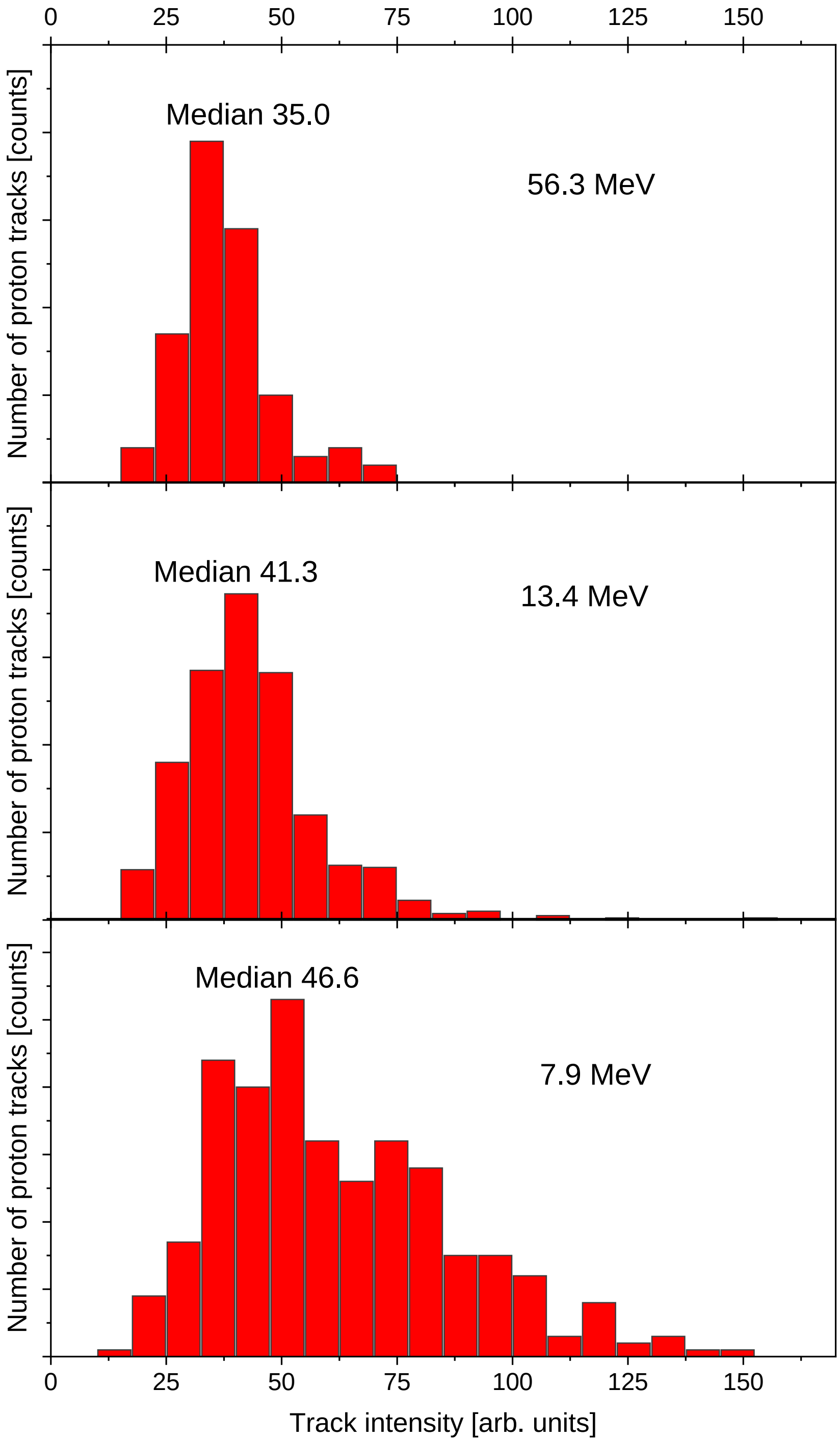

Fig.8. Distribution of track intensity for three proton energies.

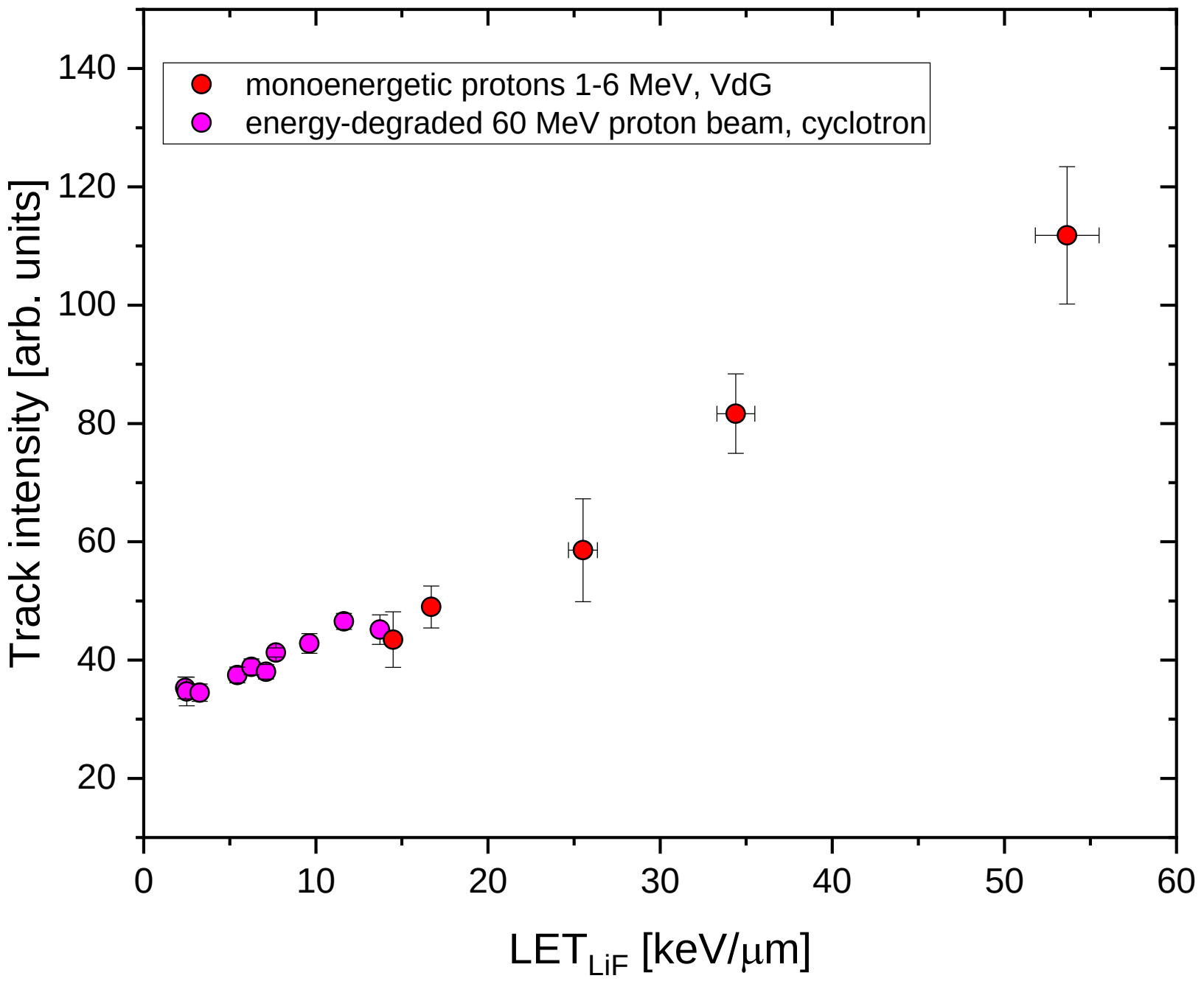

Fig. 9. Track intensity for protons vs. LET in LiF. In the case of the cyclotron irradiations, the median values of the track intensity distributions were taken.

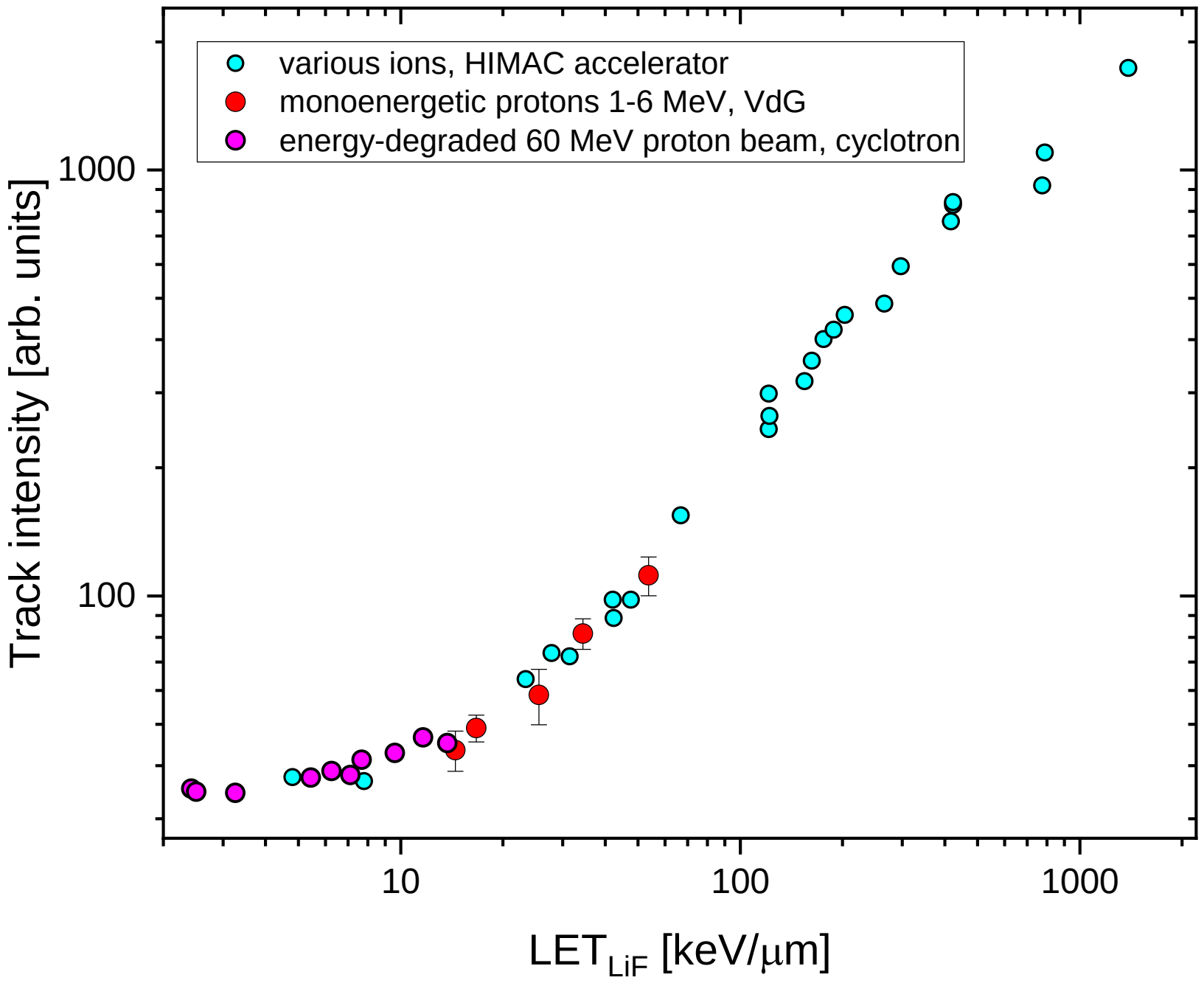

Fig. 10. Dependence of track intensity on LET for protons and various ions. In the case of the cyclotron irradiations, the median values of the track intensity distributions were taken.

It was mentioned, when discussing images presented in Fig. 6, that the number of registered tracks increases with decreasing energy of the proton beam. Quantitatively this effect is illustrated in Figs 11 and 12, which show the ratio between the number of the observed tracks and the expected number of protons impinging on the same area. This ratio strongly increases with the decreasing energy (increasing LET). For the highest studied energy only about 5% of the expected number of protons is registered in a single fluorescent image. This number increases up to 40% near the Bragg peak. The growth of the counting efficiency is caused by the shortening of the gaps between bright spots along a track and possibly also by the general brightening of the tracks, which makes weaker spots visible.

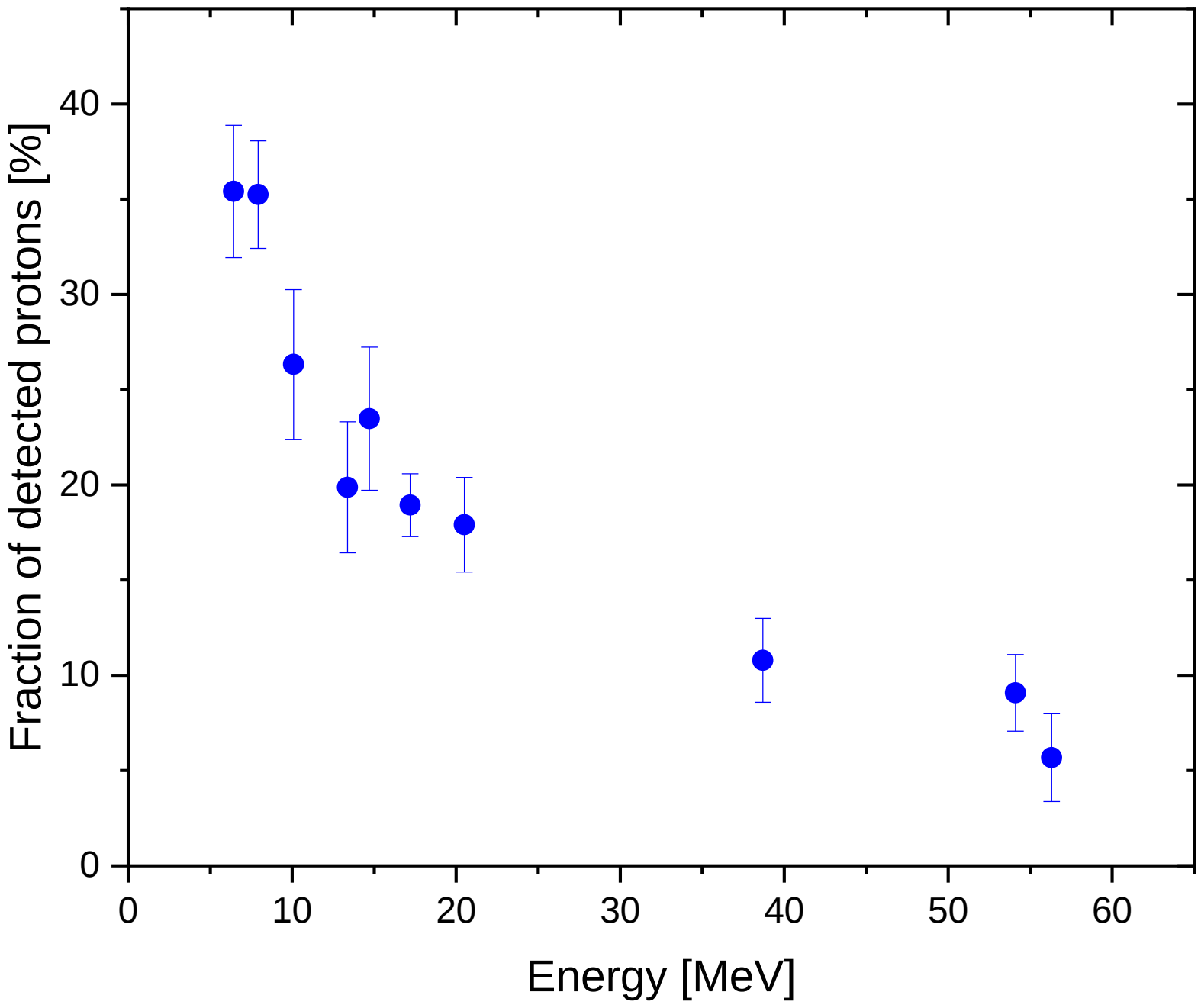

Fig.11. Ratio between the number of the registered tracks in a single slice and the proton fluence vs. proton energy.

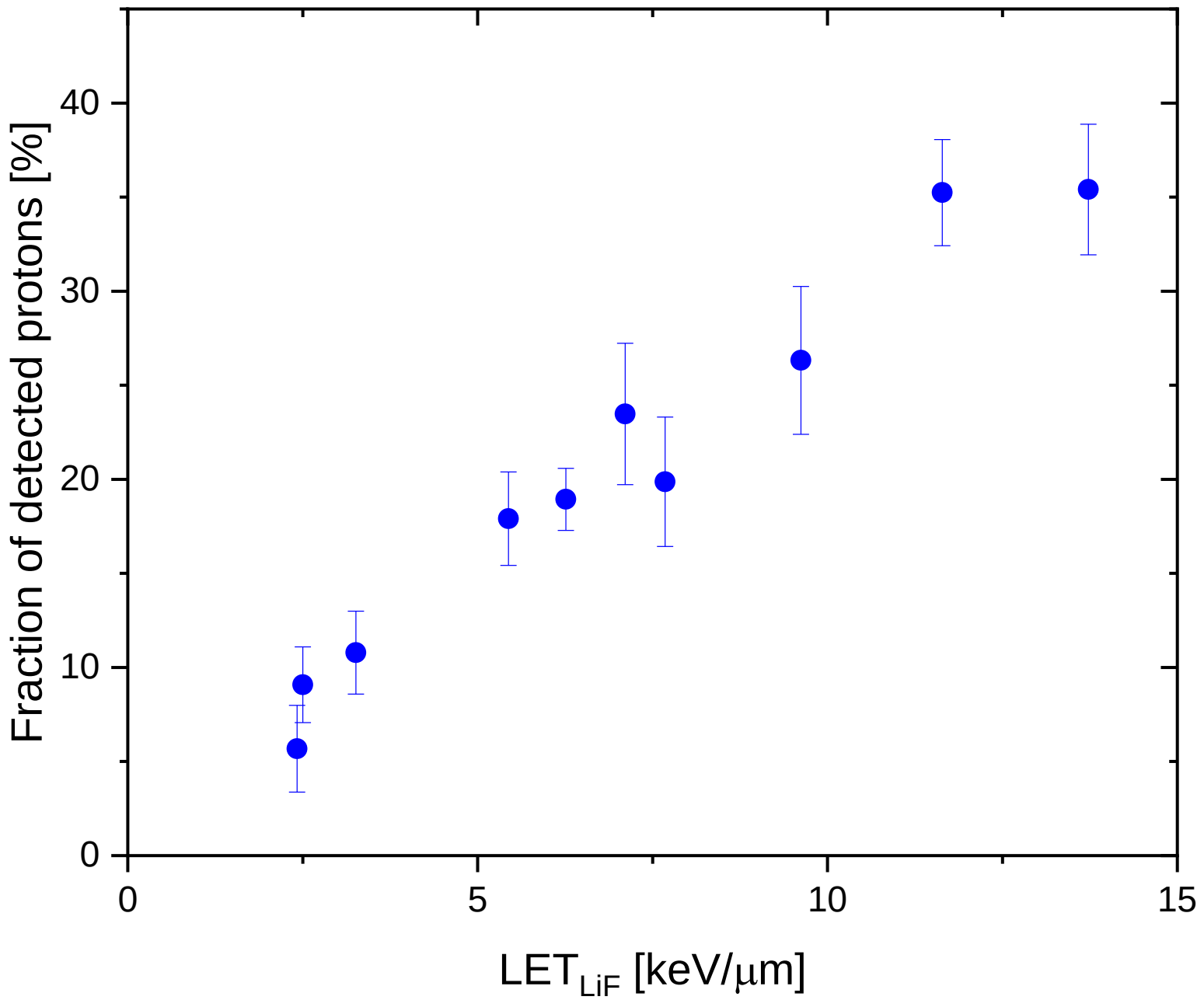

Fig. 12. Ratio between the number of the registered tracks in a single slice and the proton fluence vs. LET in LiF.

## 4. Concluding remarks

In this work, we studied the performance of LiF FNTDs after irradiations with protons with energy ranging from 1 MeV up to about 56 MeV. For all proton energies, the fluorescent tracks in the form of bright spots were observed. The distance between these spots depends on the particle energy (ionization density). For the highest of the studied energies, the spots are scattered so sparsely, that it is not possible to link spots belonging to one track. With the decreasing proton energy, the distance between spots and in consequence the number of registered spots increases. However, even for the lowest energy the number of the registered tracks is lower than the estimated number of protons per the same area. The physical mechanism causing the tracks to be not continuous seems to be related to the microscopic pattern of energy loss by a particle, but it is still not clear and this subject is currently under study.

The intensity (brightness) of the fluorescent tracks increases with the increasing LET and fits well with the trend established earlier for various ions ranging from helium to iron.

Acknowledgments

This work was funded by the National Science Centre, Poland (grant No 2020/39/B/ST9/00459). Irradiations with proton beams at Ruder Boskovic Institute were performed in the frame of the RADIATE Transnational Access project (22002881-ST-1.1-RADIATE).